\documentclass[manuscript]{acmart}

\usepackage{float}
\usepackage{supertabular}

\usepackage{tabularray}
\usepackage{caption}
\usepackage{subcaption}
\usepackage{array}
\usepackage[normalem]{ulem}
\usepackage{todonotes} 
\usepackage{soul}
\usepackage{xspace}
\usepackage{xcolor}
\usepackage{longtable}
\usepackage{tabularx}
\usepackage{graphicx}
\usepackage{enumitem,kantlipsum}
\newcommand{\nickname}{Generative Disco\xspace}
\definecolor{lightgray}{gray}{0.6}

\newcolumntype{C}[1]{>{\centering\arraybackslash}p{#1}}

\AtBeginDocument{%
  \providecommand\BibTeX{{%
    \normalfont B\kern-0.5em{\scshape i\kern-0.25em b}\kern-0.8em\TeX}}}

\setcopyright{acmcopyright}
\copyrightyear{2018}
\acmYear{2018}
\acmDOI{10.1145/1122445.1122456}

\acmConference[Woodstock '18]{Woodstock '18: ACM Symposium on Neural
  Gaze Detection}{June 03--05, 2018}{Woodstock, NY}
\acmBooktitle{Woodstock '18: ACM Symposium on Neural Gaze Detection,
  June 03--05, 2018, Woodstock, NY}
\acmPrice{15.00}
\acmISBN{978-1-4503-XXXX-X/18/06}

\begin{document}

\title{\nickname: Text-to-Video Generation for Music Visualization}


\author{Vivian Liu}
\email{vivian@cs.columbia.edu}
\affiliation{
  \institution{Columbia University}
  \city{New York}
  \state{New York}
  \country{USA}
}

\author{Tao Long}
\email{long@cs.columbia.edu}
\affiliation{
  \institution{Columbia University}
  \city{New York}
  \state{New York}
  \country{USA}
}

\author{Nathan Raw}
\email{nate@huggingface.co}
\affiliation{
  \institution{Rochester Institute of Technology}
  \city{Rochester}
  \state{New York}
  \country{USA}
}

\author{Lydia Chilton}
\email{chilton@cs.columbia.edu}
\affiliation{
  \institution{Columbia University}
  \city{New York}
  \state{New York}
  \country{USA}
}

\renewcommand{\shortauthors}{Liu et. al.}

\begin{abstract}
Visuals can enhance our experience of music, owing to the way they can amplify the emotions and messages conveyed within it. However, creating music visualization is a complex, time-consuming, and resource-intensive process. We introduce Generative Disco, a generative AI system that helps generate music visualizations with large language models and text-to-video generation. The system helps users visualize music in intervals by finding prompts to describe the images that intervals start and end on and interpolating between them to the beat of the music. We introduce design patterns for improving these generated videos: \textit{transitions}, which express shifts in color, time, subject, or style, and \textit{holds}, which help focus the video on subjects. A study with professionals showed that transitions and holds were a highly expressive framework that enabled them to build coherent visual narratives. We conclude on the generalizability of these patterns and the potential of generated video for creative professionals.



\end{abstract}





\begin{teaserfigure}
\centering
  \includegraphics[width=\textwidth]{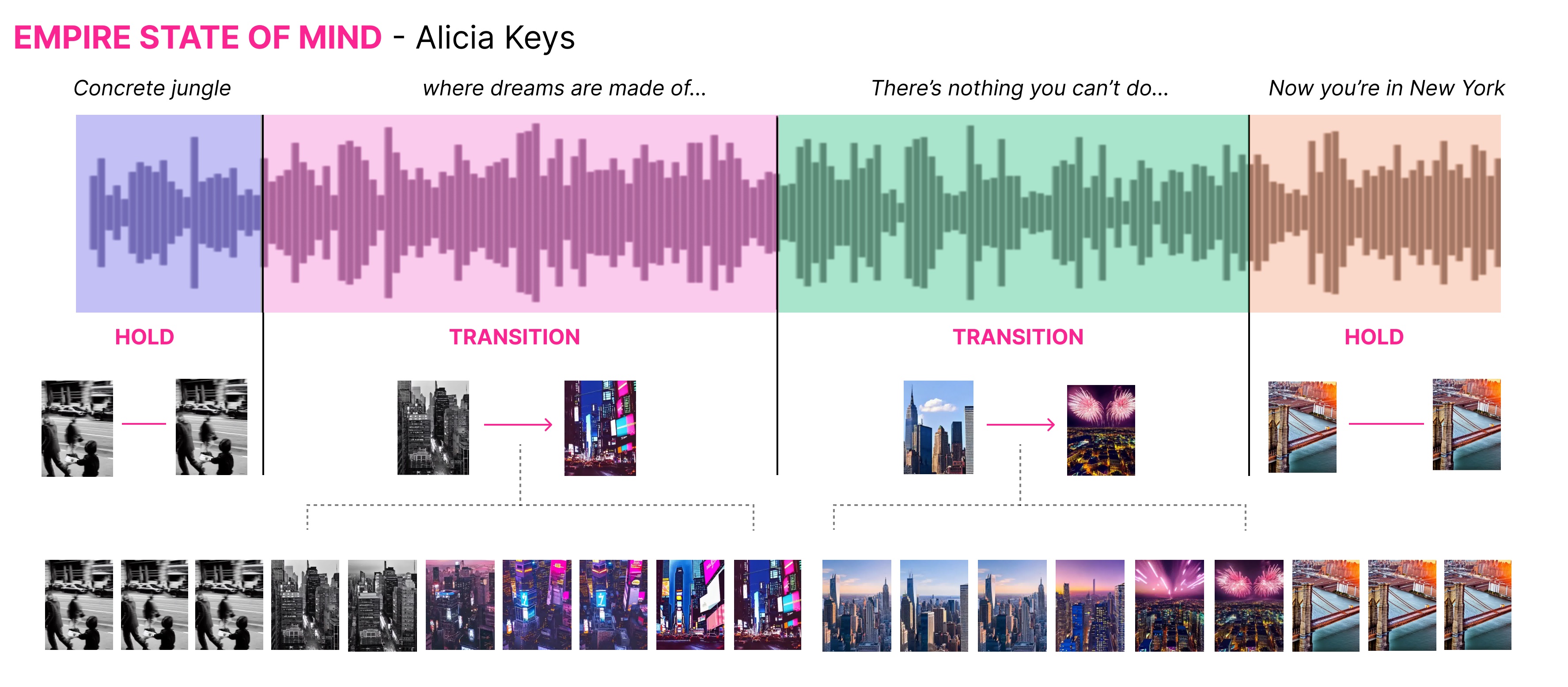}
  \caption{ \textbf{\nickname} is a text-to-video system for music visualization. Generative Disco guides users to work in intervals over a music waveform and to find generated images that define how intervals start and end. Interpolating between these images to the beat of the music creates stylized music visualization. Two design patterns, transitions and holds, guide the generative process. Transitions express change along dimensions such as color, time, subject, and style, and holds create focus on a subject.}
  \Description{The figure depicts music ("Empire State of Mind" by Alicia Keys in a waveform representation. The waveform is chunked in four colors: purple, pink, green, and orange. The purple area is the first few seconds of the song, it corresponds to the lyrics "Concrete jungle". Below it, a pair of the same two images is labeled HOLD. The images are grayscale and like stills of street photography. The next interval is longer and is highlighted in pink (corresponds to the lyrics "where dreams are made of". This area is labeled TRANSITION. Within it, a grayscale cityscape saturates to a neon image of Times Square at night. In the next green area (corresponding to the lyrics "There's nothing you can't do"), a transition occurs between a sunny cityscape and fireworks going on over a satellite view of Earth. In the last orange interval (corresponding to "Now you're in New York"), a hold focuses on a shot of the Brooklyn Bridge at sunrise.  Below the pairs of images is an expanded view that shows all the frames of the music visualization in sequence.}
  \label{fig:teaser}
\end{teaserfigure}

\maketitle

\section{Introduction}

Visuals can enhance our experience of music, because they can amplify the emotions and messages artists want to convey. In the music industry, it is standard for music to be released with music videos, lyric videos, and visualizers. Concerts and music festivals also focus on music visualization through stage displays and visual jockeying, the real-time manipulation and selection of visuals to accompany music. Music visualization has taken form in every space music can be performed, from venues \cite{vjing1,vjing2} to computer screens \cite{muvis, musicvisuals}. Visuals make music more immersive and as such, some forms of music visualization like music videos can be as beloved of a cultural production as the music itself. 



Music visualization is complex to create because the process of assembling and synchronizing visuals to music is time and resource-intensive. For example, footage for music videos has to be recorded, gathered, cut, and aligned. Every moment of the music video creation and editing process is full of artistic choices, and making these artistic choices synchronize to music is difficult. Video editors have to figure out how to make their visuals coincide with lyrics, melodies, and beats. They also have to draw upon a lot of media; often they have to scour stock footage sites, produce animations or motion graphics, or film their own footage. Collecting content can become burdensome, as users have to find video clips and assets that align with their artistic style and purchase content with limited resources.



Generative AI has the potential to help with such content creation as it can produce a wide range of customizable visuals. People are already creating music visualization using text-to-image generation, which they use to express subjects and styles that suit the vibe of their music. People post playlists to Youtube and songs on Spotify using AI-generated images as their video backgrounds and album covers \cite{abbeyroad, fotor_b}. Visuals help add an extra layer of interest to playlists and help videos get more traction on platforms by inciting click-through. Other popular trends have been to post before and after sequences of AI filters to trending sounds \cite{lensa, barbiefilter, ghiblifilter}, though these slideshows do not have the true dynamic quality of videos. 



Text-to-video models have quickly followed their text-to-image predecessors, and they introduce methods that stitch videos together by purposing text-to-image generations as video frames. These tools have the potential to produce large quantities of aesthetic content, however it is not yet fully understood how users can construct coherent and visually appealing narratives with them. The previous challenges that came with text-to-image still apply; users still have to find the right prompts, which can take exhaustive trial and error, because prompts can never fully specify images. This challenge is only furthered in the context of text-to-video. It is hard to know what to write in a prompt to reflect what video elements should change over time and what should stay the same. With generated video, it is also a challenge to mitigate motion artifacts and maintain visual consistency.

To provide a structured yet expressive approach to text-to-video generation, we introduce two design patterns: \textbf{holds} and \textbf{transitions}. \textbf{Holds} are shots where there is minimal yet subtle visual movement that allows the shot to focus on a subject. In the same way that a dancer might hold a pose for a second to put a picture to the music, holds emphasize moments within the music and give the audience visuals to anchor on. Figure 1 illustrates different holds within the song “Empire State of Mind”. Audioreactivity during holds can be minimal like this, with motion blur in the background, stars twinkling, or the color palette shifting subtly. \textbf{Transitions} are video clips that reflect visual change in terms of a dimension such as color, subject, or style. Transitions can allude to visual effects (e.g. desaturating a clip) or video filter effects (e.g. applying a watercolor filter over a clip). They can also leverage the unique properties of generative AI to create subject-to-subject transformations. Transitions help build a coherent visual narrative by implementing change through meaningful axes that people can control with prompts. In Fig. \ref{fig:teaser}, a color transition saturates as the lyrics go \textit{“Concrete jungle }[desaturated] \textit{where dreams are made of }[saturated]”. 




We present \textit{Generative Disco}, an interactive text-to-video system for music visualization. It is one of the first to explore human-computer interaction problems surrounding text-to-video systems and assist music visualization through a generative AI approach. Users can select intervals of music to visualize, which they can choose to make a transition or hold. In either case, users provide two prompts, a "start" prompt and an "end" prompt that describe the starting and ending frames of the interval. To help users explore different ways an interval could start and end, the system provides a brainstorming area that leverages a large language model (GPT-4) to provide users with prompt suggestions. These brainstorming features help users triangulate between lyrics, visuals, and music. Once users have chosen a pair of generations to start and end an interval, a music visualization clip is created by interpolating between these images to the frequency of the music. These clips taken together form the final music visualization, and we can see how ten seconds of a song can be visualized with a sequence of transitions and holds in Fig. \ref{fig:teaser}. 

To evaluate \nickname's workflow, we conducted a user study (n=12) with twelve video professionals and music experts. Our study showed that participants found the system highly expressive, easy to explore, and intuitive to use. Video professionals were able to create visuals they found useful and aesthetically pleasing while closely engaging with many facets of the music.

Our contributions are as follows:
\begin{itemize}
    \item An analysis of animated music videos to characterize design patterns for music visualization

    \item A framework for generating stylized video based on intervals as the unit of creation. Within intervals, generated video can express meaning through \textit{transitions} in color, subject, style, and time and \textit{holds} that encourage consistency and focus on a subject


    \item Generative Disco, a generative AI system that facilitates text-to-video generation for music visualization using a pipeline of a large language model, text-to-image, and text-to-video generation

    
    \item A study showing how professionals could leverage Generative Disco to express a diversity of musical genres and visual narratives
\end{itemize}

In our discussion, we elaborate use cases for how generative systems like \nickname can empower professionals and connect sound, language, and images in a workflow.

\section{Related Work}
\subsection{Music Visualization}
Music visualization can be thought of as finding visual analogues for musical elements \cite{tendulkar2020feel, hyperscore, davis18beats}. When brought into digital environments, music is often rendered as audio signals, waveforms, MIDI, lyrics, and notes. When digitized, music can be computationally analyzed in terms of its musical structure, timbre, pitch, mood, melody, harmony, dynamics (duration and volume), rhythm, and tempo. Prior works have put forward a number of approaches to synesthetically combine musical features with visual features such as color \cite{ciuha, songvis, colorbased}, shapes, line graphs, score notation, instrument visualization \cite{tunepad}, glyphs \cite{songvis, chan_classical}, and particles \cite{surveymusicvis}. For example, Davis et al. created visual analogues for rhythm and beats by temporally arranging the visible motion within a video to music to create a sense of dance \cite{davis18beats}. Various systems have also explored multimodal problems around music \cite{hyperscore, tunepad, lehtiniemi_mood} by animating instruments based on when they come in or using lyrics to find relevant pictures.


Emotions are closely intertwined with music, because music has an abstract nature that can evoke feelings. Interaction research with generative music systems have found that users like to engage with music at a high-level in terms of emotions and musical conventions (how typical or surprising music is) \cite{cococo}. Rubin and Agrawala also have shown how adding emotionally relevant segments of music to audio stories can enhance 
 storytelling \cite{underlays, rubin, rubin_editaudio}. \nickname builds on these music visualization works by analyzing music in terms of its percussive elements, dynamics (volume), and tempo and generating videos that align to these features. However, it also gives users the ability to visualize song lyrics and other higher level goals.

\subsection{Video Creation and Editing}
Because video creation and editing are time-consuming activities that have steep learning curves, there is a large body of research work behind how it can be supported and made easier. Many methods that have previously been proposed for text-based video creation relied on structured content and templates. Examples of such structured data include markdown, web pages, recipes, and knowledge graphs \cite{howtocut, url2video, videolization, recipe2video}. Text is often central to the video creation process and comes in the form of scripts, outlines, and dialogue \cite{crosspower, katika, bscript, slides_concreteness}. CrossPower is one notable example of a system that uses linguistic structures within language to layout content for animations, presentations, and videos. Video creation has also been explored from the angles of crowdsourcing \cite{vaish, bartindale}, livestreaming \cite{streamsketch}, expert patterns \cite{motif}, tutorial generation \cite{mixt, smartrecorder}, live production \cite{onbodygraphics}, and text-based exploration \cite{videodigests, sceneskim}. Other systems have explored the transformation between speech and text by leveraging text-to-speech, spoken narration, annotations, and recordings from crowdworkers \cite{crosscast, soloist, democut, quickcut, vaish, temporalseg}. 




Besides text, many systems have been designed to put music first \cite{instrumeteor,djmvp}. DJ-MVP \cite{djmvp} proposed a fully automatic system that created music videos by producing audio-video mashups from a video corpus. Other systems have been similarly reliant on source video \cite{Beatleap, DanceReproducer}. A lyrics-based system, TextAlive, automatically generated kinetic typography videos based on lyrics \cite{textalive}. Consumer-oriented software products such as CapCut \cite{capcut, premiere_autosync, doodooc} give users templates and sounds to work off of, producing short-form videos generally under a minute long. \nickname builds on this prior work by enabling faster and easier video content creation through an interactive generative AI approach centered around music.

\subsection{Generative AI in Creative Workflows}


Machine learning advances in modeling multimodal knowledge have led to meteoric improvements in generative technologies and translated into mainstream products such as Stable Diffusion \cite{stablediffusion}, Adobe Firefly \cite{firefly}, Midjourney \cite{midjourney}, and DALL-E \cite{dalle}. Early work connecting text to image mainly generated abstract and patterned images \cite{visualstyle, lemotif}. Now text-to-image tools are capable of taking in text prompts and optimizing images to capture a near infinite combination of visual concepts. Prompting, while simple conceptually, can be difficult in practice as prompts will always underspecify the image. Users may end up conducting an exhaustive trial and error search for the ideal prompt. Design guidelines, empirical studies, and papers \cite{liu2021design, initialimages, promptbook, reprompt} have shown that using subject and style keywords can help elicit high quality aesthetic outputs. 


Research systems have applied generative technologies to problems creative professionals face. Opal, a generative workflow for news illustration \cite{opal} demonstrated how LLMs can provide prompt exploration support for text-to-image generations. 3DALL-E, a text-to-image workflow that assisted product designers with conceptual CAD \cite{3dalle} demonstrated how users could integrate AI assistance at different stages of a design workflow. LLMs have also been leveraged for creative assistance in story ideation and conceptual blending to create visual metaphors, journalism angles, and science communication \cite{anglekindling, wang2023popblends,talebrush,sparks}. 

Many machine learning approaches for text-to-video generation have been suggested \cite{nuwa, phenaki, makeavideo, cogvideo}, and many text-based video AI tools and products exist \cite{WZRD, kaiber, genmo, deforum, runwaygen2}. Open-source repositories often purpose text-to-image models to convert sequences of generated images into frames for videos; Stable Diffusion Videos and Deforum \cite{deforum, discodiffusion, stablediffusionvideos} are both popular repositories that help users generate videos from sequences of prompts. Other commercial tools such as Kaiber and WZRD \cite{WZRD} help users generate by providing keyword suggestions, style presets, and video templates. Runway GEN2, another commercial tool geared towards professionals, generates high quality video clips from a single prompt--though it does not focus on audioreactivity. A design space is evidently emerging for generative systems that intertwine language, sound, and visuals. Generative Disco further opens this space and is one of the first to research human-computer interaction questions around a text-to-video approach for music visualization.  \nickname is novel in that it emphasizes more interactive control: it provides transitions and holds to punctuate and drive change throughout a music visualization and supports conceptual exploration based on the music. 



\section{Analysis of Music Visualization}

To better understand the existing design patterns and conventions behind music visualization, we conducted an analysis of popular music videos. There are many kinds of music videos; there are narrative music videos, which introduce characters and visualize explicit storylines, but there are also performance videos, which show artists performing their music with vocal or dance performance \cite{pretet2021language}. We focus on narrative videos, because we believe generative technology is best suited for artistic visuals rather than as a substitute for an artist's performance. We chose to look at animated videos in specific (as opposed to music videos featuring artists) because animated videos are a popular format where the artist possibilities are open-ended and not bounded by what can be shot in footage. Generative technology also produces stylized content that is more in line with animated videos.



\begin{figure*}[h]
    \centering
    \includegraphics[width=\textwidth]{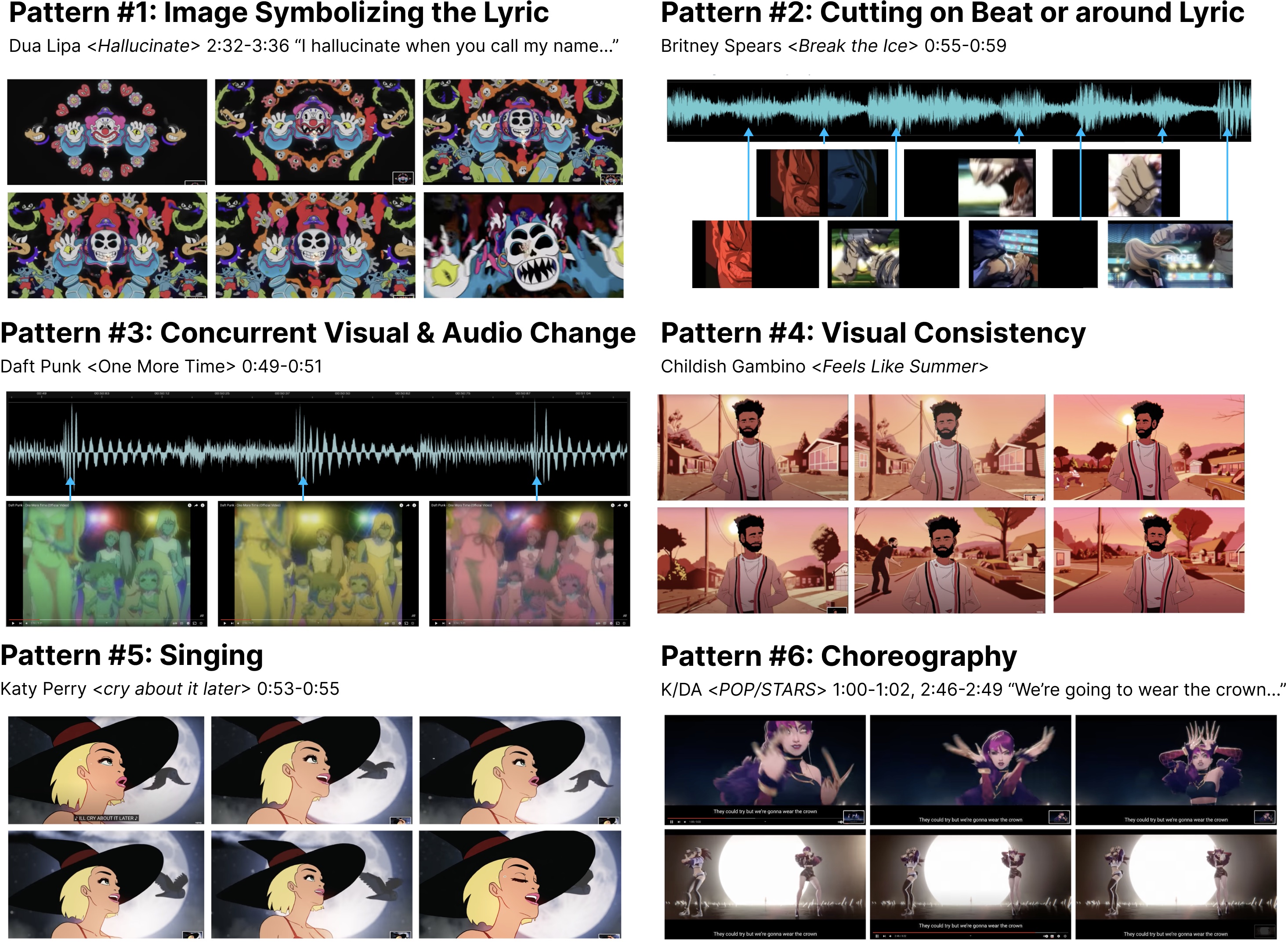}
    \caption{A list of six major audio-visual alignment patterns that we found in animated music videos. The patterns that we arrived at were 1) images symbolizing the lyric 2) cutting on the beat or around a lyric 3) concurrent visual and audio changes 4) visual consistency 5) singing 6) choreography. Examples are illustrated with popular music videos (artists labeled inline).}
    \Description{
    A 3x2 of six major audio-visual alignment patterns that we found in animated music videos. The patterns that we arrived at were 1) images symbolizing the lyric 2) cutting on the beat or around a lyric 3) concurrent video and audio changes 4) visual consistency 5) singing 6) choreography. Examples are illustrated with popular music videos such as K/DA - "Pop Stars", Dua Lipa- "Hallucinate", "Feels like Summer" - Childish Gambino, "One More Time" - Daft Punk. Arrows over the waveform representations indicate moments where the visuals align with the music --their effect is described in the main text.
    
    }
    \label{fig:mvexamples}
\end{figure*}

We collected a set of 50 animated music videos, each of which had over one million views. These videos were sourced from lists compiling popular animated music videos \cite{rollingstone, animaker, creativebloq}. This set spanned a dozen musical genres, five languages, and four decades of release dates (starting from 1985). The median view count for this dataset was 94 million views. We conducted a coding of these videos using a grounded theory approach \cite{groundedtheory}. The first and second author independently coded the videos. They have experience in professional music visualization and music analysis respectively. 

First, an open coding exercise was conducted with a small subset of these videos. Codes describing visual-specific and audio-specific changes were identified. These codes can be found within the Supplementary Material. A mapping exercise was then conducted over these audio and visual codes to create a codebook of audiovisual alignments, axially coding for patterns that would co-occur. The full set of the 50 animated music videos was then coded with this codebook. We next describe these audiovisual patterns and their frequency of appearance in the dataset.








    \textit{Images symbolizing meaning of lyrics.} 34/50 animated videos included visuals that would symbolize the lyrics. The example in Fig. \ref{fig:mvexamples} shows when the lyrics are less literal but still clearly illustrated. The lyrics go, \textit{“I hallucinate when you call my name”}, which is accompanied by a kaleidoscoping animation of skulls and clowns that grow closer to the viewer, evoking the \textit{“hallucinate”} part of the lyric. Another more literal example of this would be when a lyric goes \textit{"You're a firework"}, and fireworks explode on screen at the same time. When the lyrics and visuals align, the imagery within the music can be illustrated and the emotions can be amplified.
    
    \textit{Cutting on Beat or around Lyric.}  46/50 animated music videos would cut to the beat of the music or around a new lyric. (A cut is when the visuals would be fully replaced with new shots.) The example in Fig. \ref{fig:mvexamples} shows an animated sequence where close-ups of a face and fight scenes rapidly alternate on changes of the beat. A clear mapping between the beat structure and visuals for the listener can help them more actively engage with the structures latent in the music.

    \textit{Concurrent visual and audio changes.} 45/50 animated music videos include a significant visual change when there were drumbeats or tempo shifts. This visual change is noticeable but not as drastic as a jump cut. Instead, more subtle transitions such as color filtering, angle changes, and texture overlays are used. For example, in “One More Time” by Daft Punk, the scene color changes from green, to yellow, to red on each drumbeat (see Fig. \ref{fig:mvexamples}). These visual changes are most often quick visual transitions that help the storytelling along and establish an atmosphere.

    \textit{Visual consistency.} 40/50 animated videos included recurring shots or visuals that would reappear like visual refrains within the music. This kind of audiovisual alignment served as a visual callback to show repetition within the music (i.e. when lyrics were repeated at choruses). In Fig. \ref{fig:mvexamples}, during every instance when the artist sings \textit{“I feel like summer”}, the same visuals of the artist walking down the street would appear for a few seconds. The subtle yet recurring visuals help establish consistency within the visual narrative. As in the pattern of \textit{images symbolizing lyrics}, this audiovisual alignment helps listeners better understand the underlying structure of the music. 
    
    \textit{Singing along.} 25/50 animated videos included the main character singing along with the music. This was often straightforwardly portrayed with lipsyncing or characters at a mic.
    
    \textit{Choreography expressing the music.} 13/50 animated videos included choreography to go with music. Choreography could help physically embody a song’s emotions and lyrical content.

These identified design patterns were an actionable, useful starting point to navigate the large space of music visualization. Digital humanities scholars Shaviro and Bloombury \cite{shaviro, bloomsbury} similarly identified patterns such as rhythmic editing (which corresponds to \textit{cutting on beats and lyrics} and \textit{concurrent visual and audio changes}) and performance footage (which corresponds to \textit{singing} and \textit{choreography}). 



These are the patterns that traditional workflows can achieve. However, text-to-video models are not yet capable of reproducing them. For example, generative technologies are still not mature enough to produce singing and dancing just from text descriptions. We wanted to explore novel ways that text-to-video AI can produce dynamic visuals that symbolize music. Thus, the design of our system was also informed by their technical capabilities and what pitfalls we wanted users to avoid. For example, generated videos struggle with visual consistency. Motion artifacts are a common problem and frames within videos can seem independent of one another, yielding a jumpy, discontinuous quality. We concluded on the following design goals for our music visualization system to help users strike a balance between leveraging existing patterns and navigating the capabilities of text-to-video. 





\begin{enumerate}[leftmargin=*]


    
    \item \textbf{Holds.} Holds help users focus on a single image at specific moments within the music. This can help users express the design pattern of \textit{images symbolizing lyrics}. In holds, there is still motion and audioreactivity, though it is more subtle. One approach within the system is to encourage visual consistency by generating images from the same seed. Another approach is to generate the same prompt but with a different seed to vary the generation in aesthetic details but keep with the same subject. Holds can create helpful pauses so the video does not become a sequence of nonstop visual change. 

    \item \textbf{ Transitions.} Transitions provide visual change that corresponds with changes in the music (e.g. rising intensity within the song, changing subject matter of lyrics). Transitions paired with holds can balance consistency with variation and help users express \textit{concurrent visual and audio change}. By visualizing in a waveform where there are drops, volume changes, tempo shifts, users have awareness of where visual change can reinforce musical change. Our approach to transitions is to let users define the images that start and end them and to suggest subject and style keywords that can add meaningful differences.
    
    \item \textbf{Finding images that symbolize the music.} To help users conceptualize ways lyrics and instrumental elements can become visuals, we utilize large language models with visual knowledge (GPT-4). Prompting generative AI can easily become a search process, and helping people brainstorm can overcome this limitation.
 
    





\end{enumerate}


\section{Designing with \nickname}

\begin{figure*}[h]
    \centering
    \includegraphics[width=0.9\textwidth]{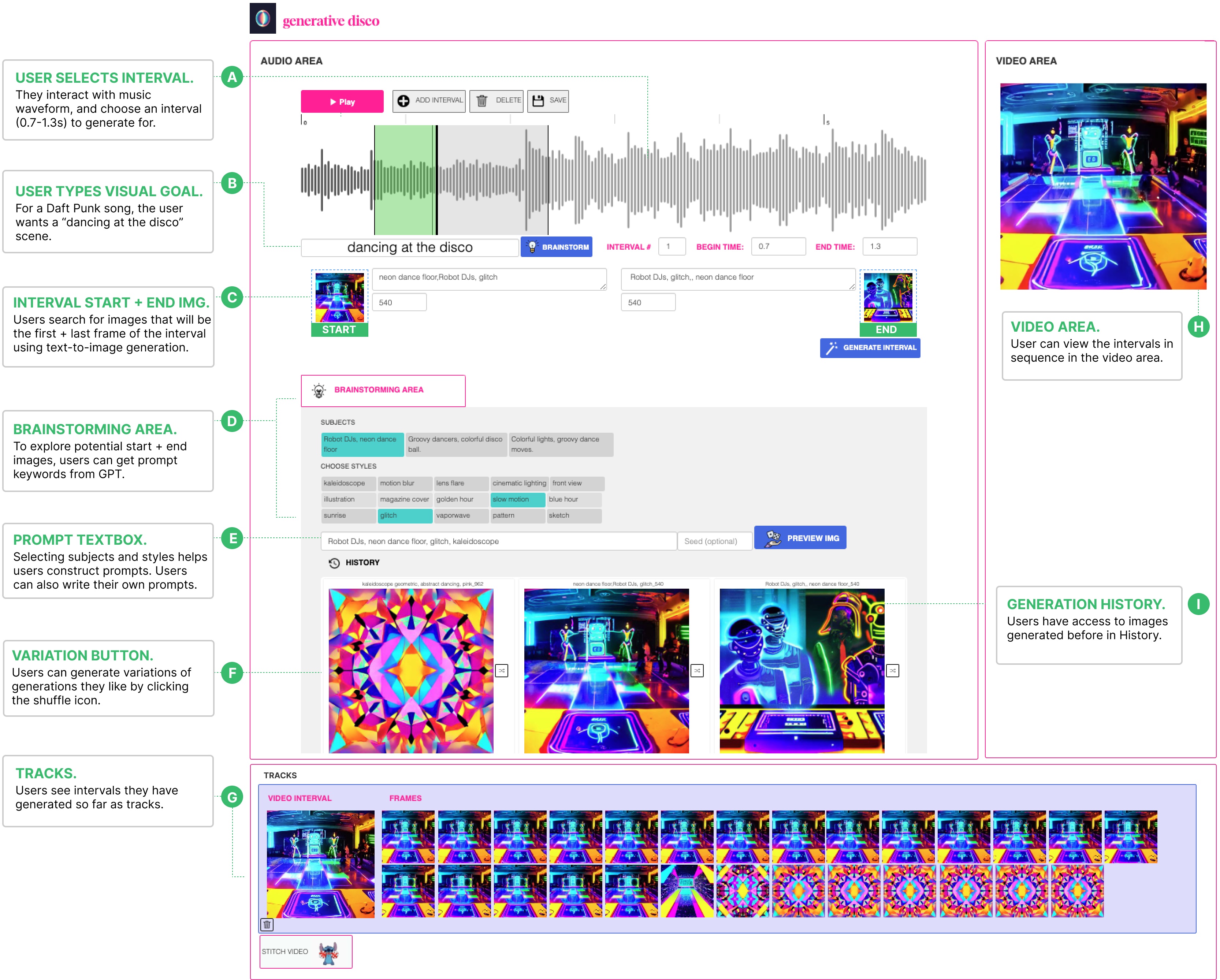}
    \caption{\nickname system design. Users begin by interacting with the waveform to create intervals within the music (A). To find prompts that will define the start and end of intervals (C), users can brainstorm prompts using suggestions from GPT-4 (B, D) and explore text-to-image generations (E, I). Results users like can be dragged and dropped into the start and end areas (C), after which an interval can be generated. Generated intervals show in the Tracks (G) and can be stitched into a video placed in the Video Area (H).}
    \Description{Generative Disco system design, annotated with captions on each side of the interface. Users begin by interacting with the waveform to create intervals within the music (top left of interface). To find prompts that will define the start and end of intervals, users can brainstorm prompts using prompt suggestions from GPT-4 or videography domain knowledge (in the BRAINSTORMING AREA, bottom left of interface) and explore text-to-image generations within the HISTORY AREA (bottom left, underneath brainstorming area). The right side of the UI contains the VIDEO AREA, for users to preview generated video. On the bottom is the TRACKS AREA, where generated video clips are shown in a frame-by-frame preview. There is a row of tiled frames showing the transition from a start prompt to an end prompt. The entire system design is grounded in an example of visualizing "One More Time" - Daft Punk using "dancing at the disco" as a high level goal. }
    \label{fig:sysdesign}
\end{figure*}

From these design goals, we designed a system for music visualization targeted at professionals. While visuals can be generated for music of any length, we focus on generating short segments of music visualization (under one minute). 

\subsection{The \nickname interface}

In the following section, we walk through how Generative Disco generates music videos. Generative Disco’s target users are audiovisual professionals and experienced hobbyists who seek novel, eye-catching visualizations. These can include professionals such as video jockeys (VJs), lyric video artists, music video editors, Youtubers and so on. We walk through a motivating example using a user persona: Naomi, who has created a remix for a Daft Punk song, "One More Time" and would like to upload her remix online. Naomi is a mashup artist who puts her remixes of pop songs on Spotify. She wants to generate a short visualizer that will loop when people listen to her remix.

The system begins by loading the audio file of her remix as a waveform (Fig.~\ref{fig:sysdesign}-A). Naomi chooses a segment she would like to visualize which has both a melodic and vocal hook. To begin working with this segment, Naomi draws intervals over her waveform by dragging around a start time and releasing over an end time. Her first interval created is three seconds long. Upon listening to it, she realizes that the melodic hook repeats 1.5 times, so she adjusts the start and end timestamps of the interval. She can do this by directly manipulating the interval boundaries or editing the BEGIN and END timestamps through text entry. 

Because Naomi would like to acknowledge the genre influences of French house and space disco that informed the song, she wants the visual to picture robots dancing at a disco party. To generate a video for a music segment, she has to define images to START and END the interval on (Fig.~\ref{fig:sysdesign}-C). To begin finding images, she begins using the BRAINSTORMING AREA (Fig.~\ref{fig:sysdesign}-D) by typing what she would like to visualize for this interval (Fig.~\ref{fig:sysdesign}-B). This description could pertain to the music (she could type in a lyric) or to a visual goal (she could describe a goal image). Naomi types \textit{"dancing at the disco"}. Upon doing this, three suggestions pop up in the BRAINSTORMING AREA (Fig.~\ref{fig:sysdesign}-D): 1)  \textit{Robot DJs, neon dance floor}. 2) \textit{ Groovy dancers, colorful disco ball}. 3) \textit{Colorful lights, groovy dance moves}. These options are suggestions for what could be the subjects for this interval and they are sourced from GPT-4.  

Naomi chooses \textit{Robot DJs, neon dance floor}, finding it relevant. The phrases are automatically copied into the prompt textbox. Below the subjects is a list of style keyword suggestions, which include keywords relevant to videography and composition. Naomi selects \textit{"glitch"} and \textit{"slow motion"}, which also automatically copy into the prompt textbox. Now with a prompt assembled (\textit{"Robot DJs, neon dance floor, glitch, slow motion"}), Naomi can hit `PREVIEW IMG' and peruse generations. Of the generations returned, Naomi likes the image featuring neon robots most heavily (right image in Fig.~\ref{fig:sysdesign}-I). To use it, she drags and drops it into START image box. This copies the relevant metadata (prompt and random initialization) into the fields next to it. 

Naomi would like there to be a scene that takes focus each time the melody repeats in the ten seconds of her remix. To make her first interval hold on that image of robot DJs on a neon dance floor, Naomi clicks on the VARIATION BUTTON (Fig.~\ref{fig:sysdesign}-F). This will give her variations of the image she chose, so that she can find a similar one to end that interval. After choosing another image from the set of variations (middle image in Fig.~\ref{fig:sysdesign}-I), she drags and drops it into the END image box, which again copies the relevant prompt metadata (Fig.~\ref{fig:sysdesign}-C). With the beginning and end prompts now set, Naomi can generate her interval and interpolate between the two images to visually fill out the interval. When the generation is complete, the interval appears in the Track Area (Fig.~\ref{fig:sysdesign}-G) as both a video and its frames.

Naomi can move onto the next interval (gray interval in Fig.~\ref{fig:sysdesign}). She has the freedom to add, edit, and delete intervals as she likes by using the buttons above the waveform (Fig.~\ref{fig:sysdesign}-A). She continues making intervals that \textit{hold} on images until she gets to the eight second mark, the moment when the vocals sing, \textit{"One more time"}. She wants dramatic visual change to coincide with the vocals, so she searches for images to create a \textit{transition}. From the BRAINSTORMING AREA, she finds the style suggestion of ``kaleidoscope" intriguing. After finding a colorful kaleidoscope pattern (left image in Fig.~\ref{fig:sysdesign}-I), she sets the new start (middle image in Fig.~\ref{fig:sysdesign}-I) and end images (right image in Fig.~\ref{fig:sysdesign}-I). This generates a transition that captures the moment the autotuned lyrics come in.

In this way, Naomi generates her music visualization interval-by-interval. We implement prompt pairs (START and END prompts) rather than prompt sequences so that users have the ability to start a new scene rather than having a nonstop sequence of images. Users can create cuts such that one image does not always flow into another. Intervals also provide users a unit abstraction to work with. When Naomi is ready, the intervals can be stitched together using the `STITCH VIDEO' button. This action places the final output video at the top right of the interface in the VIDEO AREA (Fig.~\ref{fig:sysdesign}-H).

\subsection{System Implementation}

The web application was written in Python, Javascript, and Flask. Images were generated on an NVIDIA V100. The system was built on top of two open-source repositories: a) stable-diffusion-videos \cite{stablediffusionvideos} from Hugging Face and b) wavesurfer.js \cite{wavesurfer}.


\textit{Implementing Audioreactivity}. To create audioreactive videos, we leverage existing capabilities within Stable Diffusion Videos. How this is implemented is that from an input music piece, the harmonic elements are filtered out and the percussive elements are retained. The audio is normalized and the cumulative sum of the music piece's "energy" (amplitude after normalization) is used to inform the interpolation between two images. Energy is represented as an array that starts at 0 and ends at 1; it is resized (stretched or shortened) based on the number of frames that are to be reproduced. This array determines the mix of the start and end images at each frame. Frames were collected together at a frame rate of 24 fps. The Stable Diffusion checkpoint used was V1-4. 

\textit{Interpolation.} Interpolation is controlled by a number of factors: the intensity and percussive elements of the music, the duration of the interval of music, and the frames per second. These parameters set the number of interpolation steps between the START and END images. During generation sequences, a loading image appears. Within the hardware parameters of our system, the generation of 3 images for every text prompt took approximately 10 seconds; while the generation of a 1-second interval took approximately 1 minute.


\textit{Brainstorming Suggestions.} SUBJECT suggestions are returned by prompting GPT-4: \textit{“In less than 5 words, describe an image for the following words \{DESCRIPTION\}.”} For STYLE suggestions, we compiled a list of 100 STYLE keywords. These words were curated based upon prior work that analyzed composition keywords and style exploration recommendations for AI-generated art prompts \cite{opal}. STYLE suggestions (Fig.~\ref{fig:sysdesign}-6) were retrieved by sampling from this list of words.

\textit{Generating Variations.} A variation in the context of this system is defined as an image that is varied either in terms of prompt or in terms of seed (random initialization). We return variations of images by keeping the seed constant but shuffling around comma-delimited phrases within the prompts. For example, generating the prompt \textit{"Robot DJs, neon dance floor, glitch"} would shuffle the prompt phrases for three other prompt variations, such as (\textit{"neon dance floor, glitch, Robot DJs"}, \textit{"glitch, neon dance floor, Robot DJs"}, \textit{"glitch, Robot DJs, neon dance floor")}. When the seed value is left empty, the system automatically generates the prompt with three different seeds.

\section{Evaluation}\label{sec:evaluation}
To evaluate Generative Disco as a system for music video creation, we conducted a user study that focused on the following research questions:

 \begin{itemize}
 \item \textit{RQ1. To what extent can Generative Disco help professionals produce visual narratives for music?}

 \item \textit{RQ2. What patterns of text-to-video generation emerge when users use transitions and holds to create music visualization?  } 

 \item \textit{RQ3. 
What possibilities can a generative music visualization approach like \nickname present for audiovisual professional workflows?}


\end{itemize}

\subsection{Experimental Design}

\begin{table*}[!h]
  \caption{Table of participant details including demographics, level of video experience, exposure to generative AI, and genre of music for task.}
  \Description{Table of participant details including demographics, level of video experience, exposure to generative AI, and genre of music for task.

  }
  \begin{tabular}{llllll}
    \toprule
    ID & Background    &  Video Freq  & Yrs Video  & AI-Art Freq &  Genre \\
    \midrule
P1   &    Video Professional, Lyric Videos  &  Daily & 7  & Never    &             Metalcore \\
P2   &   Video Professional, VJ  &  Daily & 14& Never  &                                     Original Composition\\
P3   &     Video Professional &   Daily &  3& Weekly &            Pop    \\
P4  &   Video Professional, live production, VJ &   Weekly &15& Weekly  &               Funk Rock \\

P5   &     Video Professional, Sound Designer &            Daily & 5 & Never  &            Alternative Indie \\
P6  &     Music Expert &  Yearly  & 4 & Yearly   &          Acoustic\\
P7   &     Music Expert, Classical + Digital &    Monthly & 0  & Never &               Hard Rock / Remix \\
P8   &       Music Expert, Acoustics + Production &              Weekly & 8 & Monthly &                                Original Composition\\
P9   &   Music Expert, Video Expert   &     Yearly & 10 &  Monthly  &                 Dance / Electronic      \\
P10   &  Video Professional, Music Videos &    Monthly & 10 &  Weekly  &            Locked Groove         \\
P11 & Video Professional, Music Videos & Daily & 6 & Weekly  & Afrobeats / Pop \\
P12   &    Music Expert   &     Yearly  & 2&    Never &                     Original Vocals / Rock \\
  \bottomrule
\end{tabular}
  \label{tab:participants}
\end{table*}

Our evaluation was conducted through a qualitative study with 12 video professionals and music experts. Our participants were recruited on a platform for freelancers where we reached out to creatives in the video editing profession and from a local organization for computer music. Participants were paid \$40 per hour for their time, and the study was conducted with each participant for 2 hours. Twelve people (6 male, 4 female, 2 non-binary) participated. The experimental IRB protocol was approved by the institution.


Participants were first interviewed about their creative expertise and about their traditional workflows for video editing. They were also asked about their level of exposure to generative AI. Many participants had exposure to generative AI and had used it professionally. ChatGPT \cite{chatgpt} was one of the most commonly cited tools by participants. Using ChatGPT, P12 had generated a royalties agreement with a client, P10 had written a routine for mixing and mastering their original music, and P3 had written video scripts for their clients. P8 and P3 had both posted their original music online with text-to-image generations as cover art. 

After a brief interview about their professional experience, an experimenter explained concepts behind Generative Disco through a slide deck that gave them a primer about text-to-image generation, prompts, and seeds. Participants were then introduced to the design patterns of transitions and holds and how it could be used to improve the quality of their AI-generated video. Afterwards, the experimenter gave a demo of the user interface by showing them how to brainstorm prompts, generate images, and choose images as start and end frames of an interval. Participants were then given a training task to generate intervals for a given song (\textit{“Empire State of Mind}” by Alicia Keys). This helped them set their expectations with the system and try the design patterns of transitions and holds.

Before the experiment, participants selected ten seconds of a song of their choice. The experimental task was to generate music visualization for that music segment over the course of one hour. After completing the open-ended experimental task, participants filled out a post-study questionnaire and were interviewed about their experience. Participant backgrounds are described in Table \ref{tab:participants}.

\section{Results}


\subsection{Quantitative Feedback}

\begin{figure*}
    \centering
    \includegraphics[width=\textwidth]{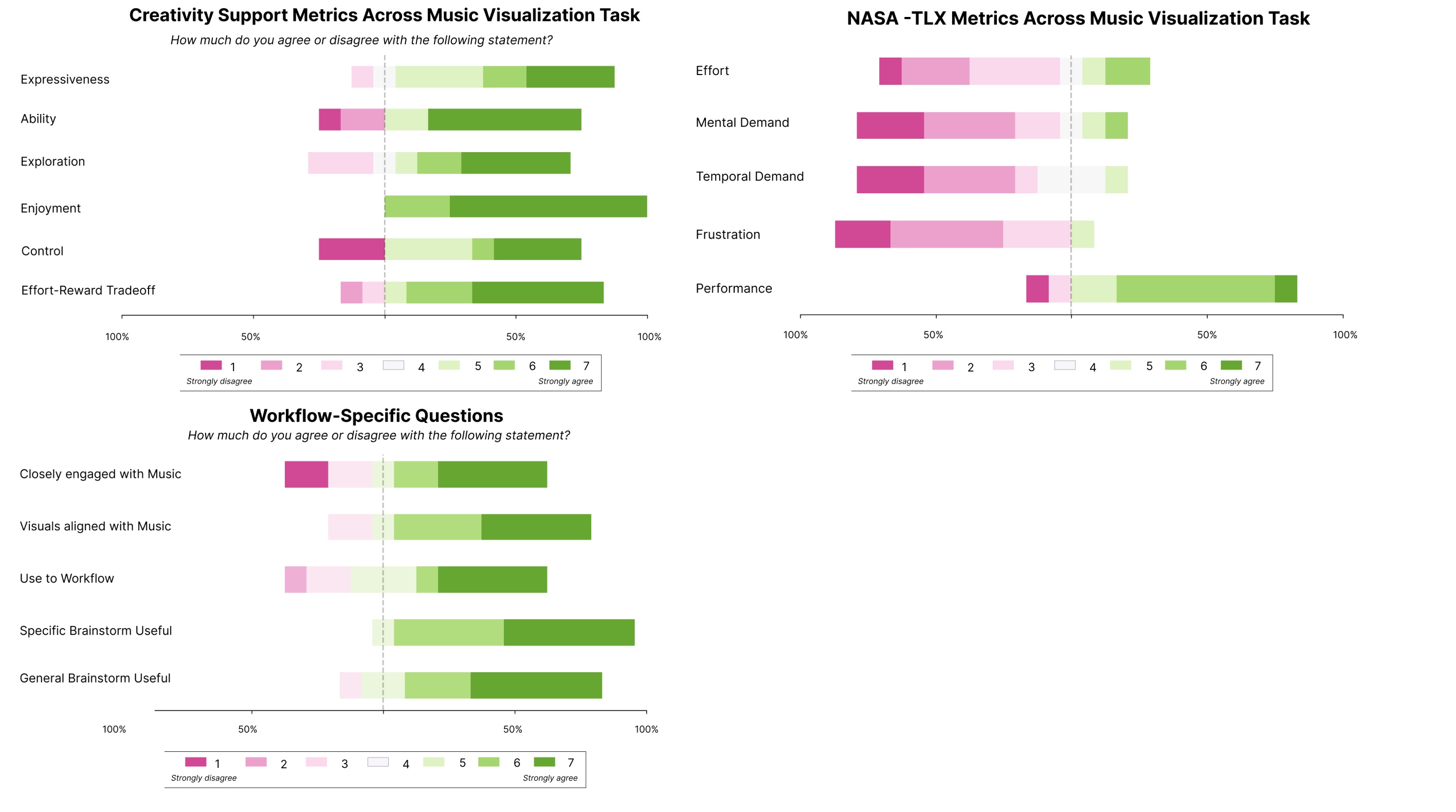}
     \caption{Distribution of Likert scale responses on NASA-TLX, creativity support index, and workflow-specific questions across all participants for the music visualization task. Full questions are in the Supplemental Material. }
    \Description{
Twelve diverging bar charts representing the distribution of Likert scale responses for NASA-TLX, CSI, and workflow-specific questions across all participants for the music visualization task. Full questions are in the Supplemental Material. Descriptions and interpretations can be found in the evaluation. Most bars on the top two categories for workflow-specific and creativity support index questions skew to the green end of the diverging color scale, symbolizing higher agreement with metrics for system performance and creativity support. For the bottom category, the bars skew pink, indicating low effort, mental load, frustration, etc. with the system.
    }
    \label{fig:quantitative}
\end{figure*}

To answer \textbf{RQ1 (``To what extent can \nickname help professionals produce visual narratives around music?")}, we report on the usefulness, usability, and creativity support of \nickname.

\subsubsection{Creativity Support Index Metrics}

\nickname performed well in terms of Creativity Support Index (CSI) \cite{csi} metrics. Responses to all questions were on a 7-point Likert scale and are visualized in the middle subplot within Figure \ref{fig:quantitative}. All twelve participants rated the system a 6 or 7 for enjoyment (median:7). Ten of 12 participants gave positive feedback (positive defined as $\geq$ 5 out of 7) that the results were worth their effort (median: 6.5). Eight of twelve participants agreed that they could sufficiently explore a number of outcomes without tedious interaction (median: 6). Expressiveness was similarly generally positive (median: 5.5). 



There was a slight split in opinion on control \textit{(``I had control over the intervals and the video I was generating''}, median: 5) Nine of twelve participants rated it 5 or higher. The remaining three participants found the system difficult to control, a problem that characterizes many generative workflows. One participant mentioned that such a system might be more difficult to use if someone has an exact picture in mind (e.g a girl singing on top of a car). There was a similar divergence of opinion for Ability \textit{(``I generated videos I would have otherwise not been able to create.'')}. Nine of twelve participants rated it highly in agreement (median: 7), while three participants gave it a 1 or 2.

\subsubsection{NASA-TLX Metrics}
The majority of participants found the performance of the system to be very positive (median: 6). The vast majority also did not find the system to be frustrating, temporally demanding (median:2), mentally demanding (median: 2), or effort-intensive (median: 3). Almost every participant (11/12) responded that their frustration during the task was low (low defined as $\leq$ 3 out of 7, median: 2). See the right side of Figure \ref{fig:quantitative}.

 \subsubsection{Workflow-Specific Questions}
 Next, we discuss participant responses to workflow-specific questions about \nickname. Participants were asked about the usefulness of \nickname to their workflow. Responses to these questions are visualized in Figure \ref{fig:quantitative}. The majority (7 of 12) rated the system positively for how closely it allowed them to engage with the music (median: 5). Nine of twelve rated the system positively for audiovisual alignment (\textit{``The system helped me come up with visuals that aligned with the music.''}, median: 5). Eleven of twelve participants were positive about the helpfulness of the specific brainstorming area, which was where GPT-4 contributed prompt suggestions (median: 5.5). Nine of twelve participants were positive that the brainstorming features was helpful, which was when sets of style keywords populated the brainstorming area (median: 5.5). When asked if \nickname would be a useful addition to their current video / music workflow, six of the participants responded with a 5 or 6 for agreement (median: 4.5).


\section{Qualitative Feedback}\label{sec:qualitative}

\subsection{Transitions and Holds as Design Patterns}
To answer \textbf{RQ2 (``What patterns of text-to-video generation emerge when users use transitions and holds to create music visualization?")} we first provide qualitative feedback from participants and later provide an analysis of the generated intervals.

\begin{figure}
    \centering
    \includegraphics[width=\textwidth]{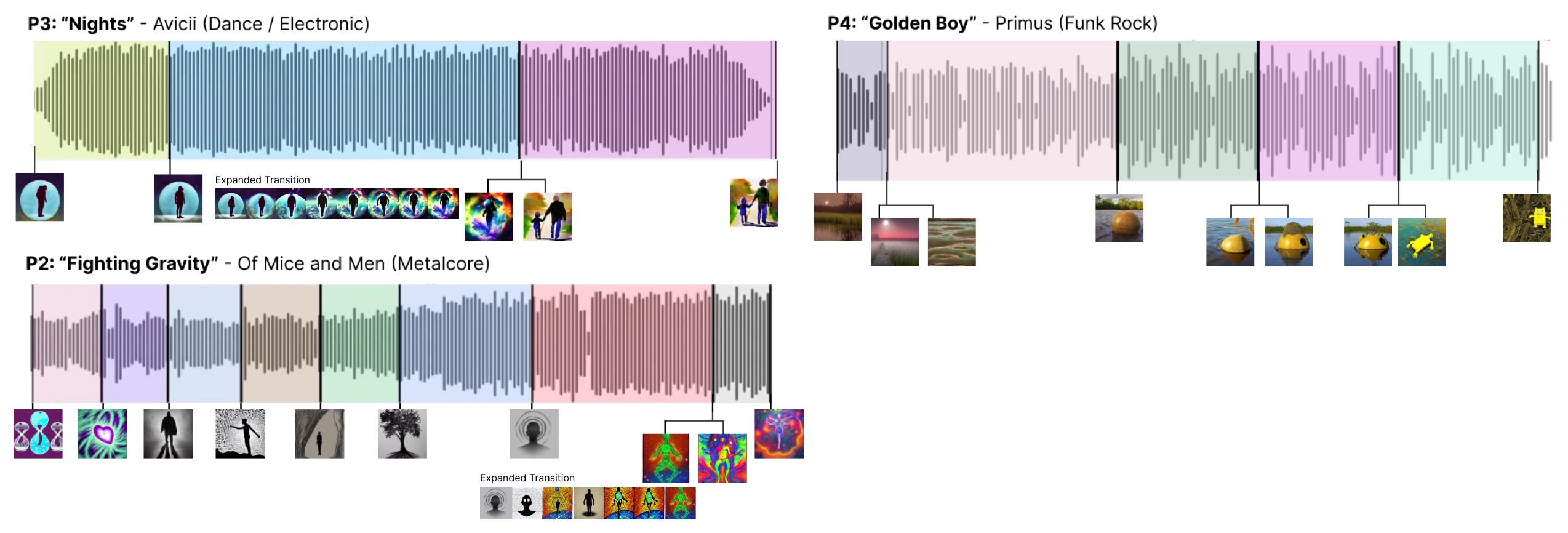}
     \caption{ Examples of different ways participants could chunk their music into intervals. P3 chunked their music around lyrics. P4 interacted with more non-lyrical elements like slides, bass, and beats. P2 captured a drop in their metalcore song. At interval boundaries, a fork between two images indicates that participants made a cut between the two images. }
    \Description{
    Three waveforms, representing music files participants worked with, are displayed. They are colored randomly. The first waveform is split in three intervals that are lengthier (3 sec each). The second waveform is split into 5 shorter ones, where the second interval is longer than the last 3 to follow. The last waveform is split into 9 short intervals that are each approximately one second long. These are examples of how participants could chunk their music in variable length intervals. P3 chunked their song around lyrics while P4 interacted with more non-lyrical elements like slides, bass, and the introduction of a beat. P1 captured a drop in a segment of an intro to a metalcore song. Annotating the start and end of the intervals are images that showed the pairs of images that were interpolated between. When two images were displayed side-by-side at the boundary of the interval, this meant that a user made a jumpcut between the two images. Two transitions are expanded and described in Section 7.1.1. 
   
    }
    \label{fig:waveform}
\end{figure}

\begin{figure}
    \centering
    \includegraphics[width=\textwidth]{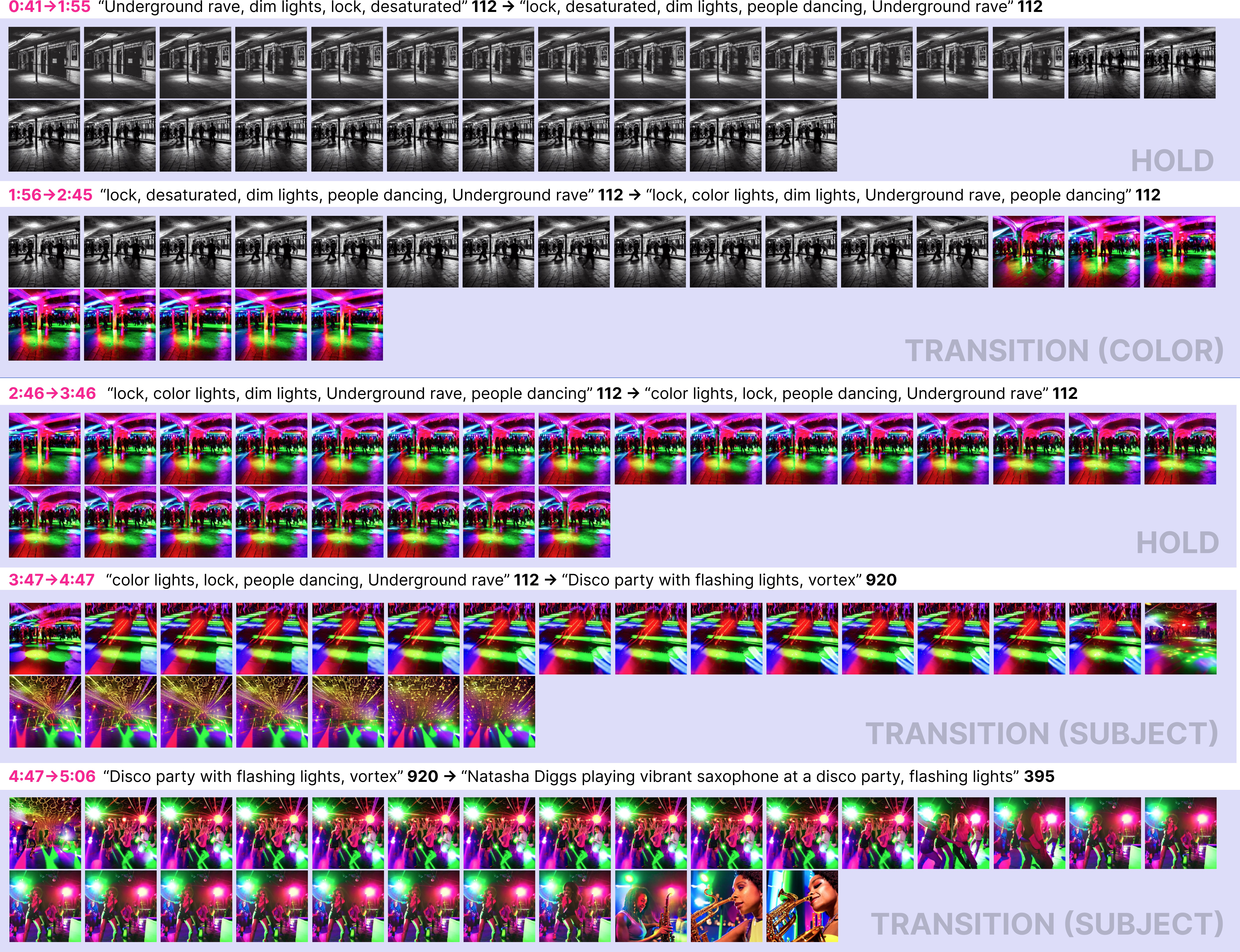}
    \caption{Frame-by-frame view of music visualization by P10. Start and end prompts are displayed above the intervals. The first and last images of each track correspond to these prompts. Intervals 1 and 3 show the design pattern of a hold. Intervals 1, 4, and 5 show color and subject transitions. }
    \Description{
    Frame-by-frame view of video created by one participant, P10 who generated a music visualization for a locked groove. Start and end prompts are displayed above the intervals. The first and last images of each track correspond to these prompts. The design pattern of a hold can be seen in Intervals 1 and 3. In these rows, the tiles that comprised each row largely stayed the same. The design pattern of a color transition can be seen in Interval 2. In these rows, the colors shifted from the start to the end, mostly going from saturation to desaturation or vice versa. Subject transitions can be seen in Intervals 4 and 5. In these rows, the following transitions occurred: "a neon party" became a "woman playing a sax", "a woman playing a sax" became "neon silhouettes", and "neon silhouettes" became "grayscale ones".
}
    \label{fig:onevideo_allframes}
\end{figure}

\subsubsection{Patterns of Engagement with Music}

Participants were able to flexibly engage with the music--the intervals they created captured a variety of elements from basic beats, vocalizations, and notes to overarching segments like lyrics and structural changes within the music (e.g. beat drops). A visual depiction of the different ways participants could chunk their music is pictured in Fig. \ref{fig:waveform}.



Five of twelve participants (P3, P12, P6, P5, P7) took a lyric-forward approach with their music. P3 chunked their music segment around lyrics. In the first and third lyric intervals (pictured in Figure \ref{fig:waveform}), the visuals hardly change--P3 \textbf{holds} on these images. In the second interval however, P3 creates a color transition from blue to rainbow to express the lyrics \textit{"live a life you will remember"} with fullness and color. Three of twelve participants interacted more with specific elements within the music, particularly when they were working with their original music (P8, P10, P11). P10 brought in a techno song with a groove that they had composed. They chunked their audio in segments based on whether it contained percussive elements or not. A frame-by-frame view of their music visualization is shown in Figure \ref{fig:onevideo_allframes}. They commented on their visual exploration in the following quote.

\begin{quotation}
\textit{``A lot of variation, especially with AI-generated music videos can make people feel crazy...[I am] trying to keep to the philosophy of what a locked groove [techno musical phrase] is, making small variations between the broader theme, from black and white to color. After four seconds, we should go from colored disco to something else... Trying to create some visuals someone in a dance euphoria would see.'' -P10}
\end{quotation}


Participants were also able to create visual changes that aligned with musical change. For example, P8 created a sharp visual change that aligned perfectly with the beginning of their melody. On the opening note of their melody, the music visualization transitioned sharply from an image of sunlight over mountains to a cabin. We see another example of audiovisual alignment when P2 generates a color transition in Figure \ref{fig:waveform} (second to last interval). This interval captured a sudden drop and the beginning of a heavy metal breakdown. After the drop, the colors became dramatic and psychedelic, reflecting the increased song intensity. 

\begin{quotation}
\textit{
``The part where there’s a slide with the bass, I wanted that to be the transition... [the system] pretty much got it… I want the intervals to start and end on beat.'' -P4}
\end{quotation}

\begin{figure}[H]
    \centering
    \includegraphics[width=\textwidth]{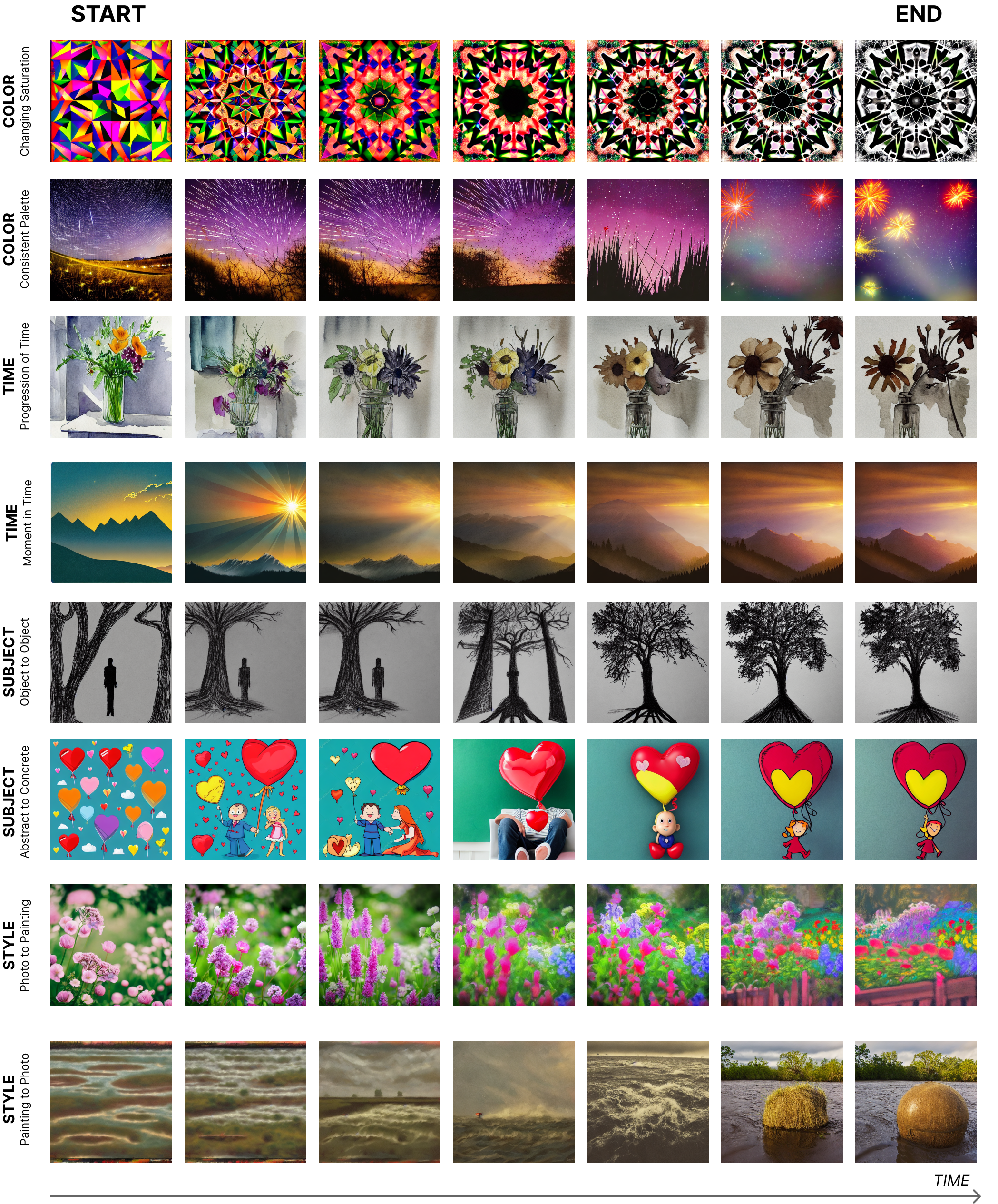}
    \caption{Illustrating transitions observed in generated intervals. Participants chose start and end prompts to transition between color palettes (rows 1-2), to establish a sense of progressing time (rows 3-4), to change from subject to subject (rows 5-6), and to add stylistic range to their generated video (rows 7-8).}
    \Description{
   In this figure we depict different types of transitions qualitatively observed during the user study: color, time, subject, and style. We show two examples for each category, where each example demonstrates a visual shift from the start image to the end image. In the first row, a colorful kaleidoscope image transitions into a grayscale one. In the second row, an image of a starry sky shifts into one of fireworks though the color scheme stays predominantly purple. In the second row, a vibrant watercolor of flowers decays into a desaturated vase, indicating the passage of time. In the fourth row, different moments of a sunrise are depicted, conveying time. In the fifth row, a man warps into a tree, showing a change in concrete subjects. In the sixth row, an animated background warps into a heart shaped character, showing another subject-to-subject transformation. In the sixth row, a photorealistic picture of flowers stylizes into a more impressionist portrayal of flowers done on pastel. In the last row an abstract, dark painting of waves shifts styles into a photorealistic look of a buoy by a swamp.
    }
    \label{fig:transitions}
\end{figure}


\subsubsection{Types of Transitions}

Throughout the video generation process, we asked participants to elaborate on how they arrived at their final intervals. We also logged all the intervals generated and performed a coding of all prompt pairs associated with intervals (60 intervals total). The first author and second author performed this coding by first conducting an open coding over the prompt pairs to identify overarching patterns for transitions and holds. This coding found that transitions could be categorized based on if they expressed change along these dimensions: color, time, subject, and style. The start and end pairs of each interval were then annotated for whether or not they contained keywords relating to transition types. For example, if color-related keywords ("red", "saturated", "neon") were found in the start and end prompts, a color transition was marked as present for that interval. We describe each type of transition as follows.

\textbf{Transition 1: Color.} Color was one of the most cited reasons behind how a participant chose the start and end images of their interval. This is reflected also in the presence of color keywords in the prompt pairs associated with each interval--40\% of the prompt pairs (Cohen's kappa = 0.46) included at least one keyword that was an explicit color term. Common color transitions participants liked to try included complete saturation changes, where the start and end image would go from grayscale to color and vice versa (P10, P5, P12, P2). Some participants also connected color more closely to musical elements; P8, for example, mapped changes in visual intensity (brightness and darkness) to changes in audio intensity. 

\begin{quotation}
    \textit{ “Songs can kind of sound brighter or darker, depending on what sound you use or the frequency range that's dominant. You can think about how color might inform our understanding of this sound, like frequency range of darker colors mapping to shorter wavelengths on a color spectrum.” - P8}
\end{quotation}

Many participants  (P7, P8, P5, P6) also used color to maintain consistency. P5 chose an image of a purple night sky and a purple sky with fireworks at the start and end of their interval based on their shared prominence of purple. This is shown in the second row of Figure \ref{fig:transitions}. P10 said that \nickname could be very useful for coming up with color corrections, as playing with color boards and scales \textit{“tends to be the hardest part of making or doing anything with videos”}.



\textbf{Transition 2: Time.} Transitions also were used to imply the progression of time (P5, P8). 20\% (Cohen's kappa = 0.46) of all prompt pairs associated with intervals contained at least one term pertaining to time. Time terms participants used included words like \textit{"timelapse"}, \textit{"speedramp"} (a VFX effect indicative of motion),  \textit{"nocturnal glow"} or pairs of time-sensitive words (transitioning from \textit{"blue hour"} to \textit{"sunset"}). P8 commented on how they tried to create a chronological sequence, \textit{“the [chosen image] seems like it's earlier in time in terms of a sunrise. It seems like chronologically it would be before the second one.”} In another example, P2 searched for two images that showed time through the decay of flowers, a transition that is pictured in row 3 of Figure \ref{fig:transitions}. P2 wanted to depict decay over time to reflect the darker undertones within their original music. 

\textbf{Transition 3: Subject.} Participants (P10, P6, P2) also liked to leverage the ability of AI to interpolate between subjects. Examples of subject-to-subject transitions can be seen in Figure \ref{fig:transitions}. Changes in the subject were identified in 26.7\% of the prompt pairs (Cohen's kappa=0.71). 


\begin{quotation}
    \textit{``I think a lot of AI-generated art has that dream-like quality. I would be curious to see a human turn into a tree. I feel like AI does that well."}-P2
\end{quotation}

P2 generated intervals that morphed a human silhouette into a tree (Fig. \ref{fig:waveform}), finding this continuous and ``dream-like" transition appealing and suitable for their song. P10 transitioned from the subject of a woman playing a saxophone to the subject of a ghost band, because they liked the idea of going from a face in sharp focus to moving bodies and wavy shapes. Another subject transition shown in Figure \ref{fig:transitions} shows an abstract subject (a pattern of hearts) transitioning into a more concrete one (a heart balloon character).

\textbf{Transition 4: Style.} Participants also liked to explore generations that started and ended with different visual styles. P2, P3, and P4 applied style transitions in their intervals, going from cartoon to photorealistic or retro to 3D. We see examples of such style transitions in the last two rows of Figure \ref{fig:transitions}. In one row, flowers shift from a photorealistic style to an oil-painted one. In the following row, the murky, "slimy" aesthetic style the participant prompted for shifts to a more photorealistic style. When the style changes are more drastic, the rate of visual change was also steeper.

\begin{figure}[b]
    \centering
    \includegraphics[width=\textwidth]{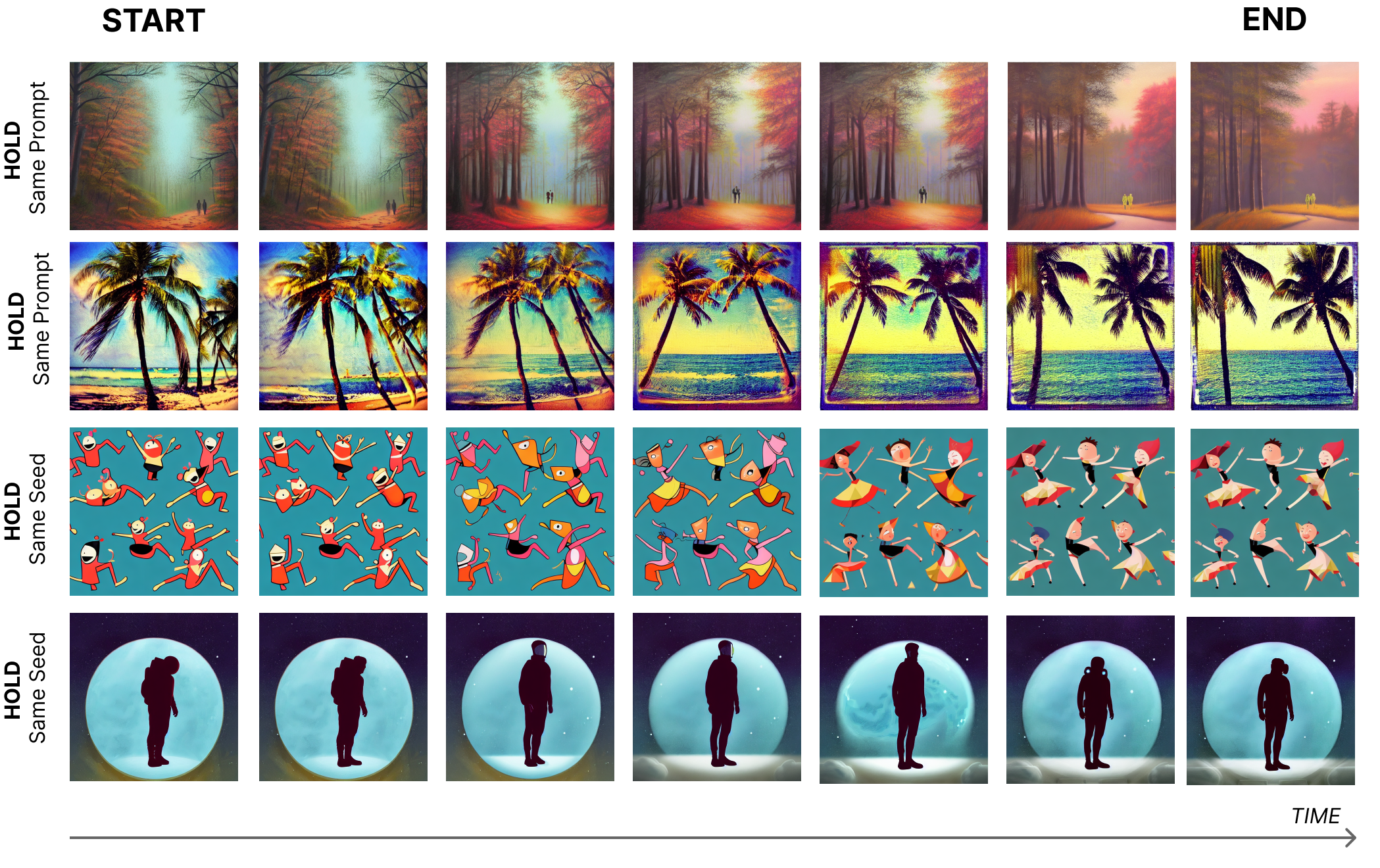}
    \caption{Illustrating holds participants created. Holds encourage focus on a subject within a generated interval. Participants achieved holds by fixing the seed or using variations of the same prompt. Examples of holds where the start and end prompt were variations of one another are seen in rows 1-2. Examples of holds where the seed is fixed are seen in rows 3-4. Holds were still dynamic. they had minimal yet subtle change (e.g. color shifts). }
    \Description{
Illustrating holds observed in generated intervals. Participants chose start and end prompts to hold and encourage consistency within their video. They generally achieved this by fixing the seed or the prompt. Examples of holds where the prompt was approximately the same are seen in rows 1-2.  In the first row, two silhouettes are maintained across a forest backdrop, even though the seed changes.  In the second row, two palm trees are in the same position and interlocked pose even though the color palette shifts across the interval. Examples of holds where the seed is fixed are seen in rows 3-4. We observe constrained changes that maintain a similar composition and color palette. In the third row, the position of six characters stay the same (preserving composition and gestalt) even though their characterization completely changes. In the fourth row, the foreground and background of an astronaut and a blue marble of earth are kept consistent even though the silhouette of the astronaut constantly shifts.
   
    }
    \label{fig:hold}
\end{figure}
\subsubsection{Holds.} Participants used the design pattern of a hold to create visual pauses within the music and to encourage consistency within the video. Participants achieved this by keeping the seed fixed or the prompts approximately the same (P10, P9, P7). 27.9\% generated intervals from images that shared the same seed. 49.1\% differed in seed but had prompts that shared the exact same set of keywords.  


We visualize holds in Figure \ref{fig:hold}. Across all rows, the composition is similar if not constant. For the first row, two figures are in a warmly tinted forest. Across the hold, the small figures persist, but the colors shift from green blue to orange red. In the last row, we see an astronaut silhouette take focus, its silhouette minimally changing across frames.

\begin{figure}[b]
    \centering
    \includegraphics[width=0.73\textwidth]{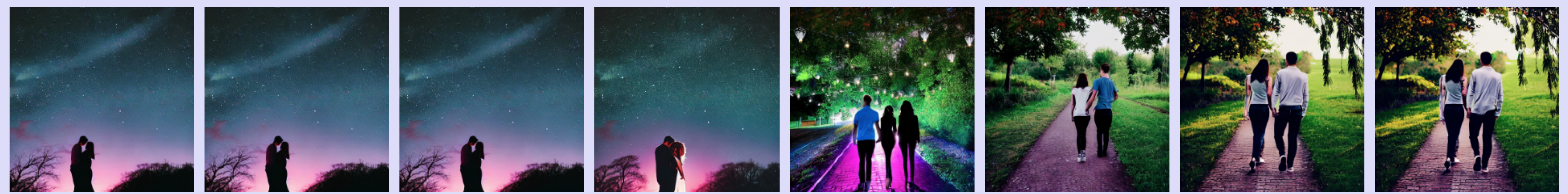}
    \caption{An example of a motion artifact from the interpolation between prompts describing a couple. In an intermediate frame, a third wheel (motion artifact) appears between the couple.}
    \Description{
   An example of a motion artifact from the interpolation between prompts "couple, soft lighting, bloom pass, street photography, back, cartoon" and "couple, soft lighting, bloom pass, street photography, back,cartoon". In an intermediate frame, instead of a couple we have a couple and a third wheel. But by the end of the interval, the image ends on just two figures, a couple.
   
    }
    \label{fig:error}
\end{figure}

\subsubsection{Visualization Pitfalls }

Consistency was one of the main deciders for what made an interval usable or not. Participants did not like it when random artifacts would appear or disappear. For example, P5 submitted two prompts about a romantic couple. However, in an intermediate frame, three people appear rather than the two (pictured in Fig. \ref{fig:error}), making it unusable for them. Whenever unwanted frames would appear for a sub-second length of time, they would introduce a glitchy, discontinuous quality.


\begin{figure}[b]
    \centering
    \includegraphics[width=0.73\textwidth]{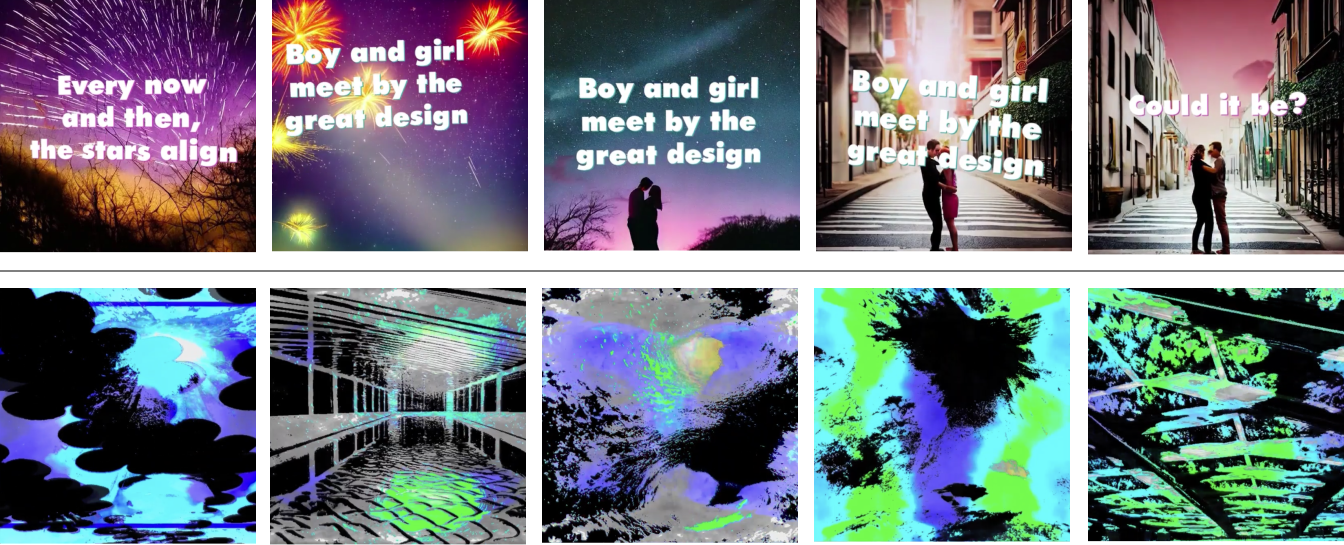}
    \caption{Stills from how participants transformed \nickname outputs with their expert workflows.
    Top:  P5 added kinetic typography to make a lyric video for Lana Del Rey's "Lucky Ones". Bottom: P9 post-processed the generated video with their signature video editing style using creative code.
   }
    \Description{ Stills from how participants transformed
    Top:  \nickname outputs with their traditional workflow. P5 added kinetic typography to make a lyric video for Lana Del Rey's "Lucky Ones". Bottom: Stills from how P9 took the generated video as video assets and post-processed them with their signature video editing style using creative code.
   
    }
    \label{fig:postprocess}
\end{figure}

The generations that were the most commonly not used to start or end intervals tended to be photorealistic. Participants (P5, P3, P1, P12, P7) did not like the distortion that could be found in bodies, faces, or hands. For example, P1 tried to generate images of a girl singing but was dissatisfied by the way her face would distort over time. P6 noted that they were not inclined to choose photorealistic generations not only because of their more apparent flaws but also because these photorealistic generations did not convey any aesthetic or mood--which they felt was the bigger picture of what music videos try to do. P5 also noted how the human figures in their images didn’t move right--they did not move like properly animated people would. To avoid distortions, a common strategy was to search for silhouettes of people or to include stylizing phrases in their prompts like \textit{“storybook illustration”},\textit{“cartoon”}, or \textit{“stylized”}. 

\subsection{Benefits of Workflow to Professionals}

We report some of the use cases and benefits professionals mentioned about \nickname, qualitatively addressing \textbf{RQ3 (``What possibilities can a generative music visualization approach like \nickname present for audiovisual professional workflows?")} Because participants spanned a broad range of video and music expertise, the use cases reported by participants for Generative Disco were different and diverse. P2, who had made over 100 lyric videos for clients in their eight years of experience, commented that \nickname would greatly benefit their workflow, which is heavily reliant on stock footage. 

\begin{quotation}
    \textit{
    “Often times looking for footage to use is very time consuming, and it's probably my least favorite part of the process ... sometimes it can get really expensive ... [This is] a whole new way of working... if I was doing it without [\nickname's] help ... it wouldn’t look as flawless the way that it does [here] with the transitions on the beat. ” -P2
    }
\end{quotation}

P5 had similar thoughts about how \nickname could assist creatives who have limited resources for footage. P2 and P3 mentioned that it could easily produce stylized video that would otherwise take them significant amounts of time and technical labor (e.g. creating a video clip filtered as if it was done in watercolors or in sketches).

\begin{quotation}
    \textit{
    “The transition from before ... was really good. I couldn't reach that with actual footage, with [stock footage site] or [stock footage site]. I would have to shoot it myself with my camera and hire a real actor... It would cost me my time, my energy. Here it was by one or two clicks."} - P5
\end{quotation}

Other professionals mentioned the potential of \nickname to provide visual assets for their work in music videos (P2) as well as live production and VJing (P4). \textit{“A lot of content for DJs doesn’t need to be real clean. It can be busy. You're providing content for a half hour, so having stuff that you don’t have to recycle, if you could have really long clips and premade stuff--that could benefit [VJs].”} They commented that these sorts of videos could be used as a mixing layer, used in the background of a main video as an asset. P10 likewise mentioned how they could use \nickname to create snippets that they could merge and layer onto real footage. P5 illustrated how they could use generated video as an animated background for a lyric video by adding kinetic typography, frames of which are shown in Fig. \ref{fig:postprocess}. 

However, participants felt that \nickname was lacking in control over camera angle compared to traditional tools. P3 and P1 wanted to be able to incorporate slow pans or zooms, but this was not supported. P1 and P4, both generative AI power users who had significant code exposure to text-to-image tools and who had used them to generate lo-fi music videos (P1) and VJ loops (P4), wanted to be able to control motion parameters such as the rotation, zoom, and translation. The desire for camera motion parameters (P2, P3, P10) emphasizes the importance of controls for these. 


\section{Discussion}

\subsection{Empowering professionals with AI}
The landscape of professional creative work is adapting. Music visualization and music videos are efforts often gated by high production costs and niches of technical expertise. In our user study, we showed how \nickname could make music visualization more accessible to music experts (P6, P7). \nickname is an instance of how generative AI can make the creative skills of different domains more accessible to one another. Many of our participants were freelancers. In a competitive global marketplace of skills, there is a pressure on freelancers to do more and to be faster. 

Being able to leverage generative mediums to work across different mediums when they have less resources can be tremendously helpful. A person posting playlists on Youtube may want a video background to make their playlist standout from the rest but may lack the artistic expertise or resources to create one themselves. Systems like \nickname can help users add visuals to their musical content so that it can better succeed on streaming and social media platforms. A video editor might have a client who wants a bespoke look that is outside of their usual set of styles. Our kind of workflow can help users easily collect personalized, stylized visuals. Instead of searching for the content, they can generate it and direct video through language. All participants, with various levels of experience with generative AI and technical background, were capable of learning how to utilize every function of \nickname within the span of fifteen minutes, which is vastly different from the steeper learning curves expected to master audiovisual tools. 



While generative tools such as Generative Disco can empower individuals to be creative, it can also introduce friction between creatives, who may have previously depended on each other rather than a tool. Further investigations can study the social dynamics surrounding generative tools within the creative industry, as done by Gero et al. in the context of LLMs for writers \cite{socialdynamics}.

\subsection{Connecting Sound, Language, and Image}

Digital humanities scholars have referred to music videos as a medium where artists often get experimental with new modes of visual expression \cite{shaviro}. As products of cutting edge tech, they are \textit{“bound to change as technologies change”}. Music videos by mainstream artists such as Linkin Park (\textit{"Lost"}) \cite{linkinpark} and Kid Cudi (\textit{"Entergalactic"}) \cite{kidcudi} have already been created with the help of generative AI. \nickname presents a generalizable approach for this emerging format of new media. 

Our results showed that \nickname is capable of creating AI-generated music visualizations that connect sound, language, and visuals. The main strength of \nickname was that it could create meaningful transitions and holds that could express both abstract and concrete notions within music. It could visualize abstract emotions and symbols but also accentuate concrete beats and specific lyrics. \nickname helped professionals explore the large space of imagery to create visual narratives that transitioned in color, time, subject, and style. Identifying these patterns can help a) better structure text-to-video user interactions b) help push generated video find its own aesthetic language, in the same way text-to-image generations have been characterized to have a certain aesthetic \cite{Midjourney_2023}. 

\nickname's workflow can also have applications and extensions into animation, motion graphics, film, and user-generated content on social media. For example, users could purpose transitions and holds to create audioreactive green screens, virtual backgrounds, and filters that align with viral sounds.

\subsection{Future Work and Limitations}

Given the duration of a user study, we could only examine \nickname in the context of 10 second video intervals. However, given more time, a user could generate videos of any length with the system. Future work could explore generating for longer durations to see how musical structure can be better alluded to (for example, repeating visuals for the chorus). Nonetheless, 10 seconds of sound is a length that people are used to engaging with, as good lengths for shortform audio content on social media can be between 15 to 45 seconds \cite{rope}. 


Another line of future work is improved motion. For example, the system could not depict a person walking with a natural walk cycle. Newer text-to-video models are starting to show signs that users can have more fine-grained control by providing depth maps, segmentation maps, and pose inputs through ControlNet \cite{controlnet}. We primarily focused on the exploration of images in an audio-first approach, but future work can try using source video as an input. Participants also noted that the interpolation set by \nickname could be at times be too audioreactive. They did not want the video to react to every change within the waveform. Some participants wanted to engage even more closely with the audio, to have user control over the intensity of the alignment to the audio, and to work with specific layers of the music such as the melody or the dynamics.



Lastly, \nickname has also been open sourced for community use. This allows us to consider running larger scale or longitudinal studies to see how artist use changes outside of a laboratory context and after extended use. 


\section{Conclusion}

Music visualization is an important and beloved cultural production that we value for the way it can make music more immersive and powerful. In this paper, we present Generative Disco, a generative AI system that helps people produce music visualizations using a large language model and text-to-video generation. These videos are generated in intervals parameterized by a start and end prompt to the frequency of the music. We introduce two design patterns for improving generated videos: transitions, which express change through color, time, subject, or style and holds, which focus on subjects and encourage consistency. In a study with audiovisual professionals, we found that \nickname could express a diversity of musical genres and visual narratives. We conclude on how workflows like Generative Disco can help creative professionals work crossmodally and empower them with more artistic reach.


\begin{acks}
\end{acks}


\bibliographystyle{ACM-Reference-Format}
\bibliography{citations2}


\end{document}


\section*{Supplemental Material for submission 2507:\\ ``Generative Disco''}

In the Formative Study, codes describing visual-specific and audio-specific changes were identified. These codes are listed below. A mapping exercise was then conducted over these audio and visual codes to create a codebook of audiovisual alignments, axially coding for patterns that would co-occur. The final codebook is detailed with visual examples in the Formative Study section.

\begin{table}[!h]
  \label{table:mv_codes}

    \centering
    \begin{tabular}{|l|}
    \hline
    \thead{Audio-Only Transitions} \\
    \hline
- Volume dynamics\\
- Tempo changes\\ 
- Percussive elements\\
- Specific background sounds \\
- Lyrics: specific lyrics words\\
- Pre-chorus (music structure)\\
- Main chorus (music structure)\\
- Bridge (music structure)\\
- Post-chorus (music structure)\\
- Drop (Trap)\\ \hline 
    \end{tabular}
      \caption{Audio-specific changes within music videos}
    \label{tab:audiocodes}
    \vspace{-6mm}
\end{table}

\begin{table}[!h]
  \label{table:mv_codes}

    \centering
    \begin{tabular}{|l|}
    \hline
    \thead{Visual-Only Transitions} \\
    \hline
- Visual intensity changes (light, saturation)\\ 
- Color filtering (color correction, color grading)\\ 
- Perspective changes (angle, camera, point of view)\\ 
- Subtle animations while the main visuals hold\\
- Full pause / hold (static visuals)\\ 
- Symbolization\\ 
- Jump cut\\ 
- Masking / composited video layers \\ 
- Text and texture overlays\\ 
- Recurring visuals and callbacks\\ 
- Choreography\\ 
- Close-ups on artist or the artist singing along
\\ \hline
    \end{tabular}
      \caption{Visual-specific changes within music videos}
    \label{tab:visualcodes}
    \vspace{-7mm}
\end{table}




\subsection*{Questions for NASA-TLX}
\begin{itemize}
    \item  \textbf{Effort.} How hard did you have to work during this task (creating a music video)?
    \item  \textbf{Mental Demand.} How mentally demanding was this task?
    \item \textbf{Temporal Demand.} How hurried or rushed was the pace of your task?
    \item \textbf{Frustration.}  How frustrated were you during the task?
    \item  \textbf{Performance.} How successful were you at accomplishing what you were asked to do?
\end{itemize}

\subsection*{Questions for Creativity Support Index}
\begin{itemize}
\item \textbf{Goal satisfaction.} How much do you agree or disagree: ``I was able to find at least one design that satisfied my goal"?
\item \textbf{Exploration.} How much do you agree or disagree: ``The system helped me fully explore the space of potential designs"?
\item \textbf{Enjoyment.} How much do you agree or disagree: ``I enjoyed working with the system."?
\item \textbf{Control.} How much do you agree or disagree: ``I felt like I had control over the generations I was creating"?  
\end{itemize}

\subsection*{Workflow-Specific Questions}
\begin{itemize}
\item \textbf{Usefulness to Workflow.} How much do you agree or disagree: ``For this task, Generative Disco would be a useful addition to my current workflow".
\item \textbf{Audio.} How much do you agree or disagree: ``I was able to closely engage with the music."
\item \textbf{Audiovisual alignment.} How much do you agree or disagree: ``For this task, the system helped me come up with visuals that aligned with the music."

\item \textbf{Feature: Specific Brainstorming.} How much do you agree or disagree: ``For this task, describing the interval and generating interval-specific brainstorming suggestions was useful for helping me come up with prompts for the music".
\item \textbf{Feature: General Brainstorming.} How much do you agree or disagree: ``For this task, the general brainstorming section (grid of prompt keywords) was useful for helping me come up with prompts".

\end{itemize}